\pdfoutput=1

\documentclass[12pt,a4paper]{article}

\usepackage{ifthen} 
\newboolean{pdflatex}
\setboolean{pdflatex}{true} 

\newboolean{articletitles}
\setboolean{articletitles}{true} 

\newboolean{uprightparticles}
\setboolean{uprightparticles}{false} 

\newboolean{inbibliography}
\setboolean{inbibliography}{false} 

\usepackage[top=1in, bottom=1.25in, left=1in, right=1in]{geometry}

\usepackage{placeins} 
\usepackage{mparhack,fixltx2e,relsize} 
\usepackage{booktabs} 

\usepackage{microtype}
\usepackage{lineno}  
\usepackage{xspace} 
\usepackage{caption} 

\usepackage{graphicx}  
\usepackage{color}
\usepackage{colortbl}
\graphicspath{{./figs/}} 

\usepackage{amsmath} 
\usepackage{amssymb}
\usepackage{amsfonts}
\usepackage{upgreek} 

\newcommand*\patchAmsMathEnvironmentForLineno[1]{%
\expandafter\let\csname old#1\expandafter\endcsname\csname #1\endcsname
\expandafter\let\csname oldend#1\expandafter\endcsname\csname
end#1\endcsname
 \renewenvironment{#1}%
   {\linenomath\csname old#1\endcsname}%
   {\csname oldend#1\endcsname\endlinenomath}%
}
\newcommand*\patchBothAmsMathEnvironmentsForLineno[1]{%
  \patchAmsMathEnvironmentForLineno{#1}%
  \patchAmsMathEnvironmentForLineno{#1*}%
}
\AtBeginDocument{%
\patchBothAmsMathEnvironmentsForLineno{equation}%
\patchBothAmsMathEnvironmentsForLineno{align}%
\patchBothAmsMathEnvironmentsForLineno{flalign}%
\patchBothAmsMathEnvironmentsForLineno{alignat}%
\patchBothAmsMathEnvironmentsForLineno{gather}%
\patchBothAmsMathEnvironmentsForLineno{multline}%
\patchBothAmsMathEnvironmentsForLineno{eqnarray}%
}

\usepackage{hyperref}    
\usepackage[all]{hypcap} 

\usepackage{xspace} 
\usepackage{upgreek}

\def\lhcb {\mbox{LHCb}\xspace}

\def\babar  {\mbox{BaBar}\xspace}
\def\belle  {\mbox{Belle}\xspace}
\def\cleo   {\mbox{CLEO}\xspace}

\def\MagUp {\mbox{\em Mag\kern -0.05em Up}\xspace}

\ifthenelse{\boolean{uprightparticles}}%
{

 \def\Ppi         {\ensuremath{\uppi}\xspace}

 \def\Pphi        {\ensuremath{\upphi}\xspace}

 \def\Ppsi        {\ensuremath{\uppsi}\xspace}

 \def\PDelta      {\ensuremath{\Delta}\xspace}                 
 \def\PXi      {\ensuremath{\Xi}\xspace}                 
 \def\PLambda      {\ensuremath{\Lambda}\xspace}                 
 \def\PSigma      {\ensuremath{\Sigma}\xspace}                 
 \def\POmega      {\ensuremath{\Omega}\xspace}                 
 \def\PUpsilon      {\ensuremath{\Upsilon}\xspace}

 \def\PB      {\ensuremath{\mathrm{B}}\xspace}                 
                  
 \def\PD      {\ensuremath{\mathrm{D}}\xspace}

 \def\PJ      {\ensuremath{\mathrm{J}}\xspace}                 
 \def\PK      {\ensuremath{\mathrm{K}}\xspace}

 \def\Pb      {\ensuremath{\mathrm{b}}\xspace}                 
 \def\Pc      {\ensuremath{\mathrm{c}}\xspace}

 \def\Ph      {\ensuremath{\mathrm{h}}\xspace}                 
 \def\Pi      {\ensuremath{\mathrm{i}}\xspace}

 \def\Pp      {\ensuremath{\mathrm{p}}\xspace}

 \def\Ps      {\ensuremath{\mathrm{s}}\xspace}

}
{

 \def\Ppi         {\ensuremath{\pi}\xspace}

 \def\Pphi        {\ensuremath{\phi}\xspace}

 \def\Ppsi        {\ensuremath{\psi}\xspace}                 
                  
 \mathchardef\PDelta="7101
 \mathchardef\PXi="7104
 \mathchardef\PLambda="7103
 \mathchardef\PSigma="7106
 \mathchardef\POmega="710A
 \mathchardef\PUpsilon="7107
                  
 \def\PB      {\ensuremath{B}\xspace}                 
                  
 \def\PD      {\ensuremath{D}\xspace}

 \def\PJ      {\ensuremath{J}\xspace}                 
 \def\PK      {\ensuremath{K}\xspace}

 \def\Pb      {\ensuremath{b}\xspace}                 
 \def\Pc      {\ensuremath{c}\xspace}

 \def\Ph      {\ensuremath{h}\xspace}                 
 \def\Pi      {\ensuremath{i}\xspace}

 \def\Pp      {\ensuremath{p}\xspace}

 \def\Ps      {\ensuremath{s}\xspace}

}

\makeatletter
\ifcase \@ptsize \relax
  \newcommand{\miniscule}{\@setfontsize\miniscule{4}{5}}
\or
  \newcommand{\miniscule}{\@setfontsize\miniscule{5}{6}}
\or
  \newcommand{\miniscule}{\@setfontsize\miniscule{5}{6}}
\fi
\makeatother

\DeclareRobustCommand{\optbar}[1]{\shortstack{{\miniscule (\rule[.5ex]{1.25em}{.18mm})}
  \\ [-.7ex] $#1$}}

\def\squark    {{\ensuremath{\Ps}}\xspace}

\def\cquark    {{\ensuremath{\Pc}}\xspace}

\def\bquark    {{\ensuremath{\Pb}}\xspace}

\def\pion   {{\ensuremath{\Ppi}}\xspace}
\def\piz    {{\ensuremath{\pion^0}}\xspace}

\def\pip    {{\ensuremath{\pion^+}}\xspace}
\def\pim    {{\ensuremath{\pion^-}}\xspace}

\def\kaon    {{\ensuremath{\PK}}\xspace}
  \def\Kbar    {{\kern 0.2em\overline{\kern -0.2em \PK}{}}\xspace}

\def\KorKbar    {\kern 0.18em\optbar{\kern -0.18em K}{}\xspace}

\def\Kp      {{\ensuremath{\kaon^+}}\xspace}
\def\Km      {{\ensuremath{\kaon^-}}\xspace}

\def\KS      {{\ensuremath{\kaon^0_{\mathrm{ \scriptscriptstyle S}}}}\xspace}

  \def\Dbar    {{\kern 0.2em\overline{\kern -0.2em \PD}{}}\xspace}
\def\D       {{\ensuremath{\PD}}\xspace}

\def\DorDbar    {\kern 0.18em\optbar{\kern -0.18em D}{}\xspace}
\def\Dz      {{\ensuremath{\D^0}}\xspace}
\def\Dzb     {{\ensuremath{\Dbar{}^0}}\xspace}

\def\Dstarp  {{\ensuremath{\D^{*+}}}\xspace}

\def\B       {{\ensuremath{\PB}}\xspace}
\def\Bbar    {{\ensuremath{\kern 0.18em\overline{\kern -0.18em \PB}{}}}\xspace}

\def\BorBbar    {\kern 0.18em\optbar{\kern -0.18em B}{}\xspace}
\def\Bz      {{\ensuremath{\B^0}}\xspace}

\def\Bu      {{\ensuremath{\B^+}}\xspace}

\def\Bp      {{\ensuremath{\Bu}}\xspace}

\def\Bd      {{\ensuremath{\B^0}}\xspace}
\def\Bs      {{\ensuremath{\B^0_\squark}}\xspace}

\def\Bdb     {{\ensuremath{\Bbar{}^0}}\xspace}
\def\Bc      {{\ensuremath{\B_\cquark^+}}\xspace}
\def\Bcp     {{\ensuremath{\B_\cquark^+}}\xspace}

\def\jpsi     {{\ensuremath{{\PJ\mskip -3mu/\mskip -2mu\Ppsi\mskip 2mu}}}\xspace}

  \def\Y#1S{\ensuremath{\PUpsilon{(#1S)}}\xspace}

\def\proton      {{\ensuremath{\Pp}}\xspace}
\def\antiproton  {{\ensuremath{\overline \proton}}\xspace}

\def\Lz          {{\ensuremath{\PLambda}}\xspace}
\def\Lbar        {{\ensuremath{\kern 0.1em\overline{\kern -0.1em\PLambda}}}\xspace}
\def\LorLbar    {\kern 0.18em\optbar{\kern -0.18em \PLambda}{}\xspace}

\def\Lb      {{\ensuremath{\Lz^0_\bquark}}\xspace}

\def\Lc      {{\ensuremath{\Lz^+_\cquark}}\xspace}

\def\BF         {{\ensuremath{\mathcal{B}}}\xspace}

\newcommand{\decay}[2]{\ensuremath{#1\!\to #2}\xspace}         

\def\to                 {\ensuremath{\rightarrow}\xspace}

\def\CP                {{\ensuremath{C\!P}}\xspace}

\def\AT#1     {\ensuremath{A_{\mathrm{T}}^{#1}}\xspace}           

\def\C#1      {\ensuremath{\mathcal{C}_{#1}}\xspace}                       
\def\Cp#1     {\ensuremath{\mathcal{C}_{#1}^{'}}\xspace}                    
\def\Ceff#1   {\ensuremath{\mathcal{C}_{#1}^{\mathrm{(eff)}}}\xspace}        
\def\Cpeff#1  {\ensuremath{\mathcal{C}_{#1}^{'\mathrm{(eff)}}}\xspace}       
\def\Ope#1    {\ensuremath{\mathcal{O}_{#1}}\xspace}                       
\def\Opep#1   {\ensuremath{\mathcal{O}_{#1}^{'}}\xspace}                    

\newcommand{\tev}{\ifthenelse{\boolean{inbibliography}}{\ensuremath{~T\kern -0.05em eV}}{\ensuremath{\mathrm{\,Te\kern -0.1em V}}}\xspace}
\newcommand{\gev}{\ensuremath{\mathrm{\,Ge\kern -0.1em V}}\xspace}
\newcommand{\mev}{\ensuremath{\mathrm{\,Me\kern -0.1em V}}\xspace}
\newcommand{\kev}{\ensuremath{\mathrm{\,ke\kern -0.1em V}}\xspace}
\newcommand{\ev}{\ensuremath{\mathrm{\,e\kern -0.1em V}}\xspace}
\newcommand{\gevc}{\ensuremath{{\mathrm{\,Ge\kern -0.1em V\!/}c}}\xspace}
\newcommand{\mevc}{\ensuremath{{\mathrm{\,Me\kern -0.1em V\!/}c}}\xspace}
\newcommand{\gevcc}{\ensuremath{{\mathrm{\,Ge\kern -0.1em V\!/}c^2}}\xspace}
\newcommand{\gevgevcccc}{\ensuremath{{\mathrm{\,Ge\kern -0.1em V^2\!/}c^4}}\xspace}
\newcommand{\mevcc}{\ensuremath{{\mathrm{\,Me\kern -0.1em V\!/}c^2}}\xspace}

\def\invfb   {\ensuremath{\mbox{\,fb}^{-1}}\xspace}

\newcommand{\chisq}{\ensuremath{\chi^2}\xspace}

\newcommand{\chisqip}{\ensuremath{\chi^2_{\text{IP}}}\xspace}

\def\gsim{{~\raise.15em\hbox{$>$}\kern-.85em
          \lower.35em\hbox{$\sim$}~}\xspace}
\def\lsim{{~\raise.15em\hbox{$<$}\kern-.85em
          \lower.35em\hbox{$\sim$}~}\xspace}

\def\sPlot{\mbox{\em sPlot}\xspace}
\def\sWeight{\mbox{\em sWeight}\xspace}

\def\ptot       {\mbox{$p$}\xspace}
\def\pt         {\mbox{$p_{\mathrm{ T}}$}\xspace}

\def\tell1  {TELL1\xspace}
\def\ukl1   {UKL1\xspace}

\newcommand{\eg}{\mbox{\itshape e.g.}\xspace}

\def\Bds{{\ensuremath{\B^0_{\kern -0.1em{\scriptscriptstyle (}\kern -0.05em\squark\kern -0.03em{\scriptscriptstyle )}}}}\xspace}

\renewcommand{\epsilon}{\varepsilon}
\renewcommand{\theta}{\vartheta}

\newcommand{\ppKK}{\proton\antiproton\kaon\kaon}
\newcommand{\ppKPi}{\proton\antiproton\kaon\pion}
\newcommand{\ppPiPi}{\proton\antiproton\pion\pion}
\newcommand{\pphh}{\ensuremath{\proton\antiproton h \xspace
h^{\prime}\xspace}}
\newcommand{\phh}{\ensuremath{\proton h \xspace h^{\prime}\xspace}}
\newcommand{\pp}{\proton\antiproton}
\newcommand{\hh}{\ensuremath{ h \xspace h^{\prime}\xspace}}

\newcommand{\KPi}{\kaon\pion}

\def\BTopphh{\decay{\Bds}{\pphh}}

\def\BToppKK{\decay{\Bd}{\ppKK}}
\def\BsToppKK{\decay{\Bs}{\ppKK}}

\def\BToppKPi{\decay{\Bd}{\ppKPi}}
\def\BsToppKPi{\decay{\Bs}{\ppKPi}}

\def\BToppPiPi{\decay{\Bd}{\ppPiPi}}
\def\BsToppPiPi{\decay{\Bs}{\ppPiPi}}

\def\sBToJPsiKst  {\decay{\Bd}{\jpsi K^*(892)^0}}
\def\JPsiTopp  {\decay{\jpsi}{\proton\antiproton}}

\def\BdToppKst  {\decay{\Bd}{\proton\antiproton K^*(892)^0}}

\def\hadronp{{\ensuremath{\Ph^+}}\xspace}
\def\hadronm{{\ensuremath{\Ph^-}}\xspace}
\def\LbTophhh{\decay{\Lb}{\proton\hadronm\hadronp\hadronm}}

\newcommand{\BuPPbarK}{\texorpdfstring{\decay{\Bu}{\proton \antiproton
\Kp}}{}}

\usepackage{cite} 
\usepackage{mciteplus}

\begin{document}

\renewcommand{\thefootnote}{\fnsymbol{footnote}}
\setcounter{footnote}{1}

\begin{titlepage}
\pagenumbering{roman}

\vspace*{-1.5cm}
\centerline{\large EUROPEAN ORGANIZATION FOR NUCLEAR RESEARCH (CERN)}
\vspace*{1.5cm}
\noindent
\begin{tabular*}{\linewidth}{lc@{\extracolsep{\fill}}r@{\extracolsep{0pt}}}
\ifthenelse{\boolean{pdflatex}}
{\vspace*{-2.7cm}\mbox{\!\!\!\includegraphics[width=.14\textwidth]{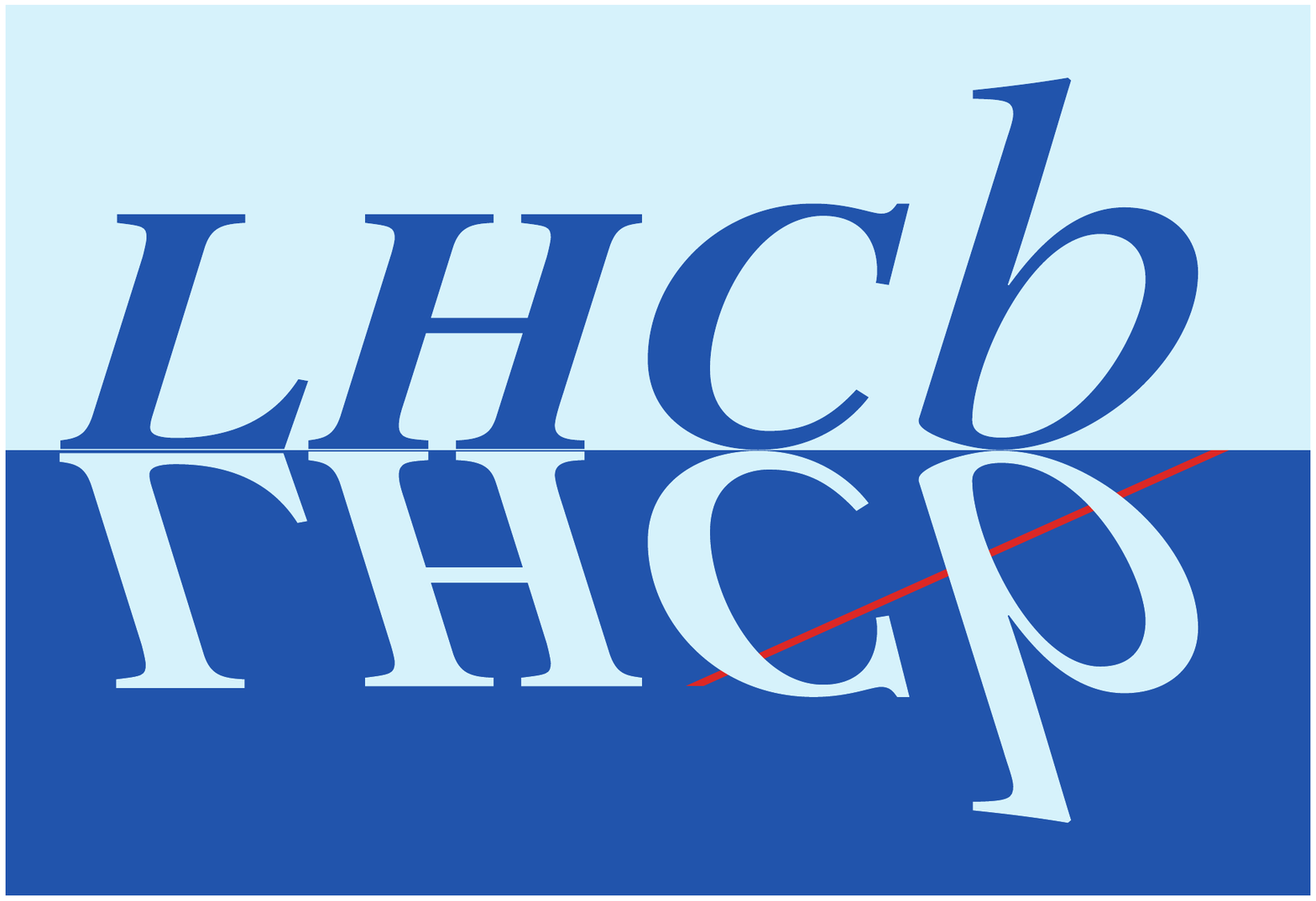}} & &}%
{\vspace*{-1.2cm}\mbox{\!\!\!\includegraphics[width=.12\textwidth]{lhcb-logo.eps}} & &}%
\\
 & & CERN-EP-2017-052 \\  
 & & LHCb-PAPER-2017-005 \\  
 & & September 27, 2017 \\ 
 & & \\
\end{tabular*}

\vspace*{3.5cm}

{\normalfont\bfseries\boldmath\huge
\begin{center}
Observation of charmless baryonic\\
decays $\B^0_{\kern -0.1em{\scriptscriptstyle (}\kern -0.05em\squark\kern -0.03em{\scriptscriptstyle)}} \to \proton \antiproton h^+ h^{\prime-}$
\end{center}
}

\vspace*{2.0cm}

\begin{center}
The LHCb collaboration\footnote{Authors are listed at the end of this article.}
\end{center}

\vspace{\fill}

\begin{abstract}
  \noindent
Decays of $B^0$ and $B_{s}^0$ mesons to the charmless baryonic final
states $p \overline{p} h^+ h^{\prime-}$, where $h$ and
$h^\prime$ each denote a kaon or a pion,
are searched for using the LHCb detector. The analysis is based on a sample of
proton-proton collision data collected at center-of-mass energies of
$7$ and $8\,$TeV, 
corresponding to an integrated luminosity of $3\,$fb$^{-1}$.
Four-body charmless baryonic \Bs decays are observed for the first time.
The decays $B^0_{s}\to p \overline{p} K^+ K^-$, $B^0_{s}\to p
\overline{p} K^\pm \pi^\mp$, 
$B^0\to p \overline{p} K^\pm \pi^\mp$ and
 $B^0\to p \overline{p} \pi^+\pi^-$ 
are observed with a significance greater than $5$ standard deviations;
evidence at $4.1$ standard deviations
is found for the $B^0\to p \overline{p} K^+ K^-$ decay and an upper limit is
set on the branching fraction for $B^0_{s}\to p \overline{p} \pi^+ \pi^-$.
Branching fractions in the kinematic region $m(p
\overline{p})<2850\,$MeV$/c^2$ are measured relative to the 
$B^0 \to J/\psi (\to p \overline{p}) K^*(892)^0$
 channel. 
\end{abstract}

\vspace*{2.0cm}

\begin{center}
  Published in Phys.~Rev.~D~96~(2017)~051103(R)
\end{center}

\vspace{\fill}

{\footnotesize 
\centerline{\copyright~CERN on behalf of the \lhcb collaboration, licence \href{http://creativecommons.org/licenses/by/4.0/}{CC-BY-4.0}.}}
\vspace*{2mm}

\end{titlepage}


\newpage
\setcounter{page}{2}
\mbox{~}

\cleardoublepage

\renewcommand{\thefootnote}{\arabic{footnote}}
\setcounter{footnote}{0}



\pagestyle{plain} 
\setcounter{page}{1}
\pagenumbering{arabic}

\newpage

\noindent
In recent years, studies by the \lhcb collaboration have greatly increased the knowledge of the decays
of \B mesons to final states containing baryons.
The first observation of a baryonic \Bc decay was reported in
2014~\cite{LHCb-PAPER-2014-039}, and \lhcb recently reported the first observation
of a baryonic \Bs decay~\cite{LHCb-PAPER-2017-012}, the last of the four \B meson
species for which a baryonic decay mode had yet to be observed.

Primary areas of interest in baryonic $B$ decays include the hierarchy of branching fractions to
the various decay modes, the presence of resonances and the existence of a threshold enhancement
in the baryon-antibaryon mass spectrum~\cite{Hou:2000bz,Bevan:2014iga}.
First evidence of \CP violation in baryonic $B$ decays has been reported
from an analysis of \BuPPbarK decays~\cite{LHCb-PAPER-2014-034}.
It is of great interest to search for further manifestations of \CP violation in baryonic
\B decays, \eg with so-called triple-product correlations (TPCs),
see Ref.~\cite{Geng:2008ps} and references therein.
For certain decays asymmetries of up to $20\%$ are predicted~\cite{Geng:2006jt}.
Four-body decays are particularly suited for this approach since the
definition of the TPCs
 do not involve the spins of the
final-state particles, unlike TPCs in three-body decays~\cite{Gronau:2011cf,Geng:2008ps}.

This paper presents a search for the decays of \Bz and \Bs mesons to the
four-body charmless baryonic final states $\proton \antiproton h^+ h^{\prime-}$,
where $h$ and
$h^\prime$ each denote a kaon or a pion.
The inclusion of charge-conjugate processes is implied, unless otherwise indicated.
For simplicity, the charges of the $h^+ h^{\prime-}$ combinations will be omitted
unless necessary. 
The branching fractions of these baryonic decays are
measured relative to the
$B^0 \to J/\psi (\to p \overline{p}) K^*(892)^0$ channel. 
So far only the resonant decay $\BdToppKst$ has been
seen by the \babar~\cite{Aubert:2007qea} and
\belle~\cite{Chen:2008jy} collaborations,
which measured its branching fraction to be $\mathcal{B}(\BdToppKst) =
(1.24^{+0.28}_{-0.25})\times 10^{-6}$~\cite{PDG2016}.
 An upper limit $\mathcal{B}(\Bd\to\pp\pip\pim) < 2.5 \times 10^{-4}$ at
 90\% confidence level 
has been set by the \cleo collaboration~\cite{Bebek:1988jy}.

The data sample analyzed corresponds to an integrated luminosity of
1\invfb of proton-proton collision data at a center-of-mass energy
of 7\tev and 2\invfb at 8\tev.
The LHCb detector~\cite{Alves:2008zz,LHCb-DP-2014-002} is a
single-arm forward spectrometer covering the pseudorapidity range $2 < \eta < 5$, designed for
the study of particles containing $b$ or $c$ quarks. The detector elements that are particularly
relevant to this analysis are as follows: a silicon-strip vertex detector surrounding the proton-proton interaction
region that allows $c$ and $b$ hadrons to be identified from their characteristically long
flight distance; a tracking system that provides a measurement of momentum, $p$, of charged
particles; two ring-imaging Cherenkov detectors that are able to discriminate between
different species of charged hadrons; and calorimeter and muon systems
for the measurement of photons and
neutral hadrons, and the detection of penetrating charged particles.
 Simulated data samples, produced with software described 
in Refs.~\cite{Sjostrand:2006za,*Sjostrand:2007gs,LHCb-PROC-2010-056,Lange:2001uf,Golonka:2005pn,Allison:2006ve,*Agostinelli:2002hh,LHCb-PROC-2011-006}, 
are used to evaluate the response of the detector and to
investigate possible sources of background.

Real-time event selection is performed by a
trigger~\cite{LHCb-DP-2012-004} that consists of a hardware stage, 
based on information from the calorimeter and muon systems, followed
by a software stage, which performs a full event reconstruction.
The hardware trigger stage requires events to have a muon with high
transverse momentum, \pt, or a
hadron, photon or electron with high transverse energy in the
calorimeters.
Signal candidates may come from events where the hardware trigger was caused either by signal particles
or by other particles in the event. 
The software trigger requires a two-, three- or four-track
secondary vertex with a significant displacement from any primary
proton-proton interaction vertices (PVs). At least one charged particle
must have $\pt > 1.6\gevc$ and be inconsistent with originating from a PV.
A multivariate algorithm~\cite{BBDT} is used for
the identification of secondary vertices consistent with the decay
of a \bquark hadron.

The final selection of \Bds candidates, formed by combining 
four
charged hadron candidates 
-- a proton, an antiproton and an oppositely charged pair of light mesons --
is carried out with a
filtering stage followed by  requirements on the response of a
boosted decision tree (BDT) classifier~\cite{Breiman,AdaBoost} and on particle identification (PID). 
The filtering stage includes
requirements on the quality, \ptot, \pt and \chisqip of the
tracks, loose PID requirements and an upper limit on
the $\pp$ invariant mass;
the \chisqip is defined as the difference between the vertex-fit \chisq of a PV reconstructed
with and without the track in question.
Each \Bds candidate must have a good-quality vertex that is displaced from the
associated PV (that with which it forms the smallest \chisqip), must
satisfy \ptot and \pt requirements, 
and must have a reconstructed invariant mass close to
that of a \Bds meson under the signal mass hypothesis.
A requirement is also imposed on the angle $\theta_{\rm dir}$ between
the candidate momentum vector and the line between the associated PV
and the candidate decay vertex.

There are 15 input quantities to the BDT classifier: \pt, $\eta$, \chisqip,
$\theta_{\rm dir}$ and the flight distance of the \Bds candidate;
the quality of the \Bds vertex fit; the \pt and \chisqip of the
tracks;
and the largest distance of closest approach between any pair of
 tracks.
The BDT is trained using simulated $\BTopphh$ signal candidates, generated with
uniform distributions over phase space, and events in a high sideband of
the $\ppKPi$ invariant mass in data ($m(\ppKPi)$ in the range
5450--5550\mevcc) 
to represent the background.
Tight PID requirements are applied to all final-state particles to reduce the combinatorial
background, suppress the cross-feed backgrounds between the different $\pphh$
final states --- background from other signal decays
where one particle is misidentified ---
and ensure that 
the datasets for the three $\pphh$ final states are mutually exclusive.
For each final state individually, the requirements on the PID and BDT
response are optimized for the signal significance using simulation samples for the signal.
After all selection requirements are applied, approximately 3\% of
events with at least one candidate also contain a second candidate; 
a candidate is then selected at random. The efficiency of the full
reconstruction and selection, including the acceptance and the trigger selection, is
approximately $0.1\%$.

To reject contributions from intermediate charm states,
candidates with $\hh$ invariant mass consistent with a \Dz meson or $\phh$
invariant mass consistent with a \Lc baryon are removed.
The contribution from the charmonium region is removed by requiring the
invariant mass of the $\pp$ pair to be less than 2850\mevcc,
similar to the procedure in Refs.~\cite{LHCb-PAPER-2016-001,LHCb-PAPER-2014-034}.
This last requirement is not applied to the
normalization mode \mbox{$\sBToJPsiKst$}, where the vector mesons are
reconstructed in the \mbox{\decay{\jpsi}{\proton\antiproton}} and
\mbox{\decay{K^*(892)^0}{\Kp\pim}} decay modes.
All the other steps of the selection
are in common for the signal and the normalization modes.

The yields of the signal decays are obtained from a simultaneous unbinned
extended maximum likelihood  fit to the $\Bds$ candidate 
invariant mass distributions in the three $\pphh$
final states in the range 5165--5525\mevcc.  
This approach accounts for potential cross-feed from one channel to another due to
particle misidentification.
Each signal component is modeled with a double-sided Crystal Ball (DSCB)
function~\cite{Skwarnicki:1986xj}.
For each signal the tail parameters of the DSCB functions are determined from simulation.
The peak position of the \Bd signals is common to the three final states,
while the difference between the peak positions of the \Bd and \Bs
signals is constrained to its known value~\cite{PDG2016}.
The width of the \Bd signal is a free parameter in the $\ppKPi$ final state
and it is related to the width in the other two final states
by scale factors determined from simulation.
The same applies to the width of the \Bs signals, which is a free
parameter only
in the $\ppKK$ final state.

For each final state the dominant $\BTopphh$ cross-feed background
is included: the $\BToppKPi$ mode in the $\ppKK$ and $\ppPiPi$ invariant mass
distributions, and the $\BToppPiPi$ mode in the $\ppKPi$ spectrum.
Each cross-feed background is modeled with a DSCB function with all
the shape parameters fixed according to simulation;
 the yield is fixed
relative to the yield in the correctly reconstructed final state
taking into account the (mis)identification probabilities calibrated
using data, as described below.
In addition, a combinatorial background component modeled by an
exponential function, with both parameters free to vary, is
present for each final state.

The yield of the normalization decay is determined from a separate simultaneous fit to the
$\ppKPi$, $\pp$ and $\KPi$ invariant mass distributions in the ranges
5180--5380\mevcc, 3047--3147\mevcc and 642--1092\mevcc, respectively. The $\sBToJPsiKst$ component is
parameterized in the $\KPi$ invariant mass
distribution by a relativistic spin-1 Breit-Wigner function and
in the $\ppKPi$ and $\pp$ invariant mass distributions by DSCB functions with
the tail parameters fixed from simulation. The $\KPi$
S-wave component is modeled in the $\KPi$ invariant mass distribution 
by the LASS parametrization~\cite{Aston:1987ir,Aubert:2008zza}
that
describes nonresonant and $K^*_0(1430)^0$ S-wave contributions;
this component is modeled in the $\ppKPi$ and $\pp$ invariant mass distributions with
the same shape as the $\sBToJPsiKst$ component. A combinatorial background component
modeled by a freely varying exponential function is also present in each spectrum.

\begin{figure*}[!tb]
  \centering
  \includegraphics[width=0.49\textwidth]{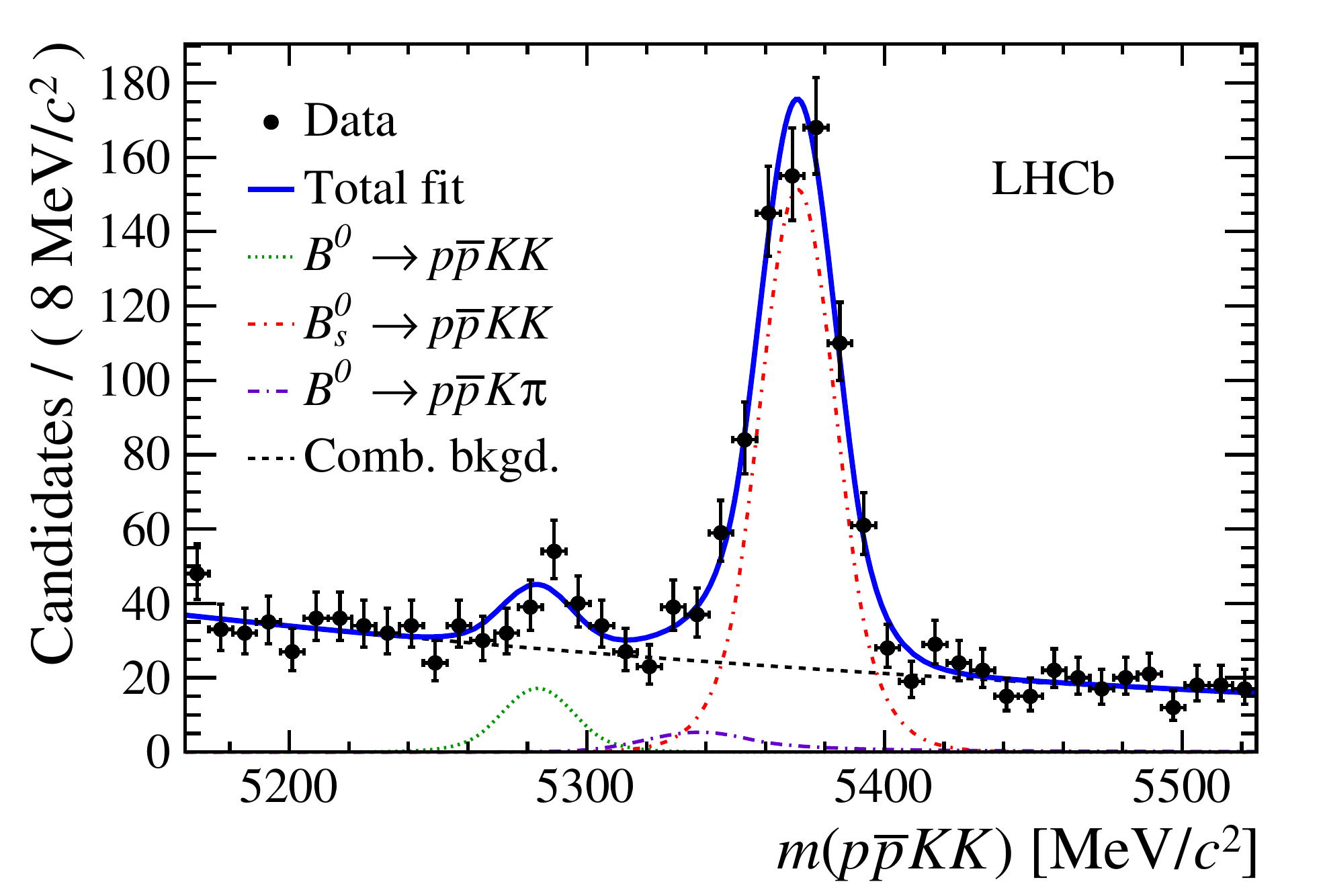}
  \includegraphics[width=0.49\textwidth]{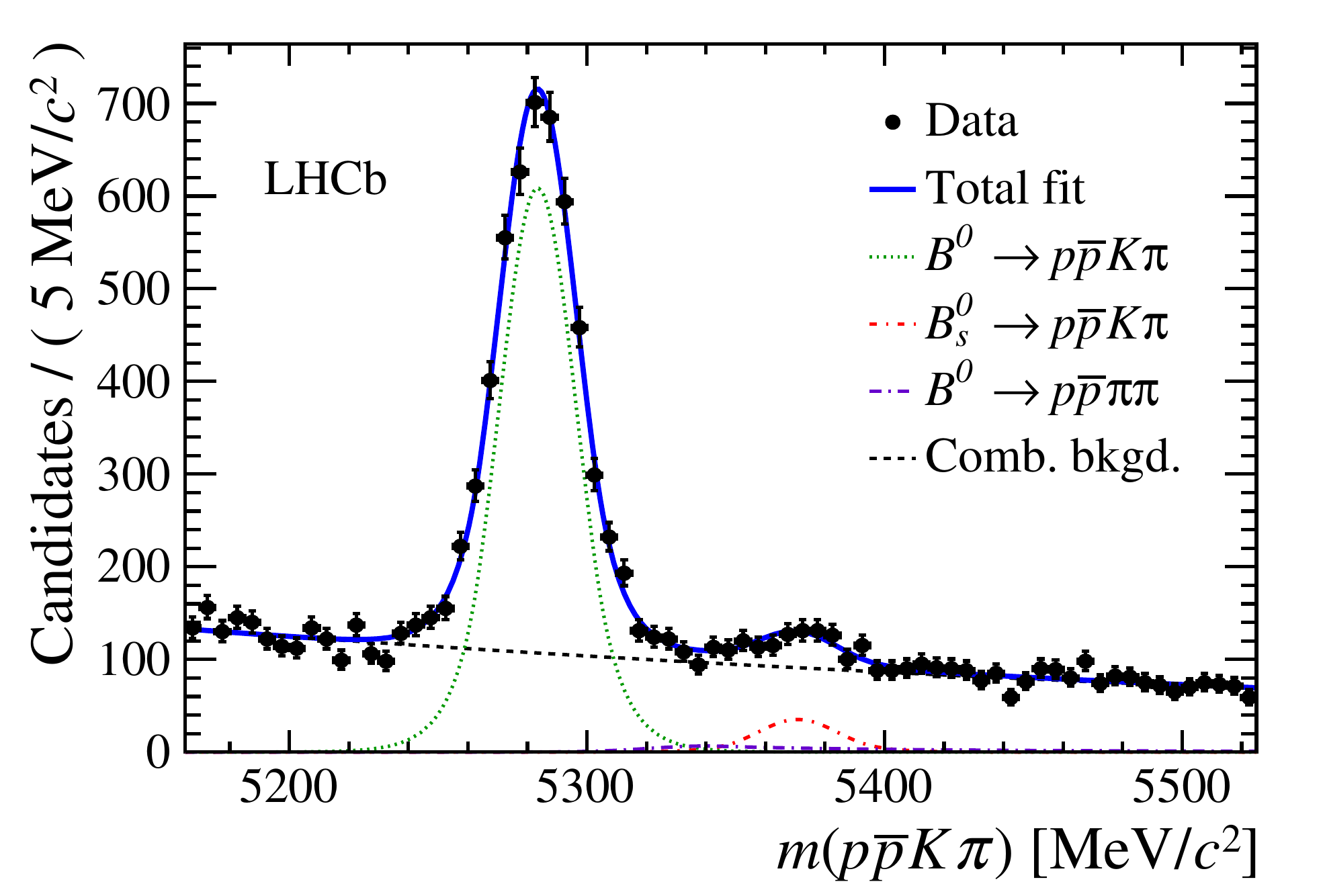}
  \includegraphics[width=0.49\textwidth]{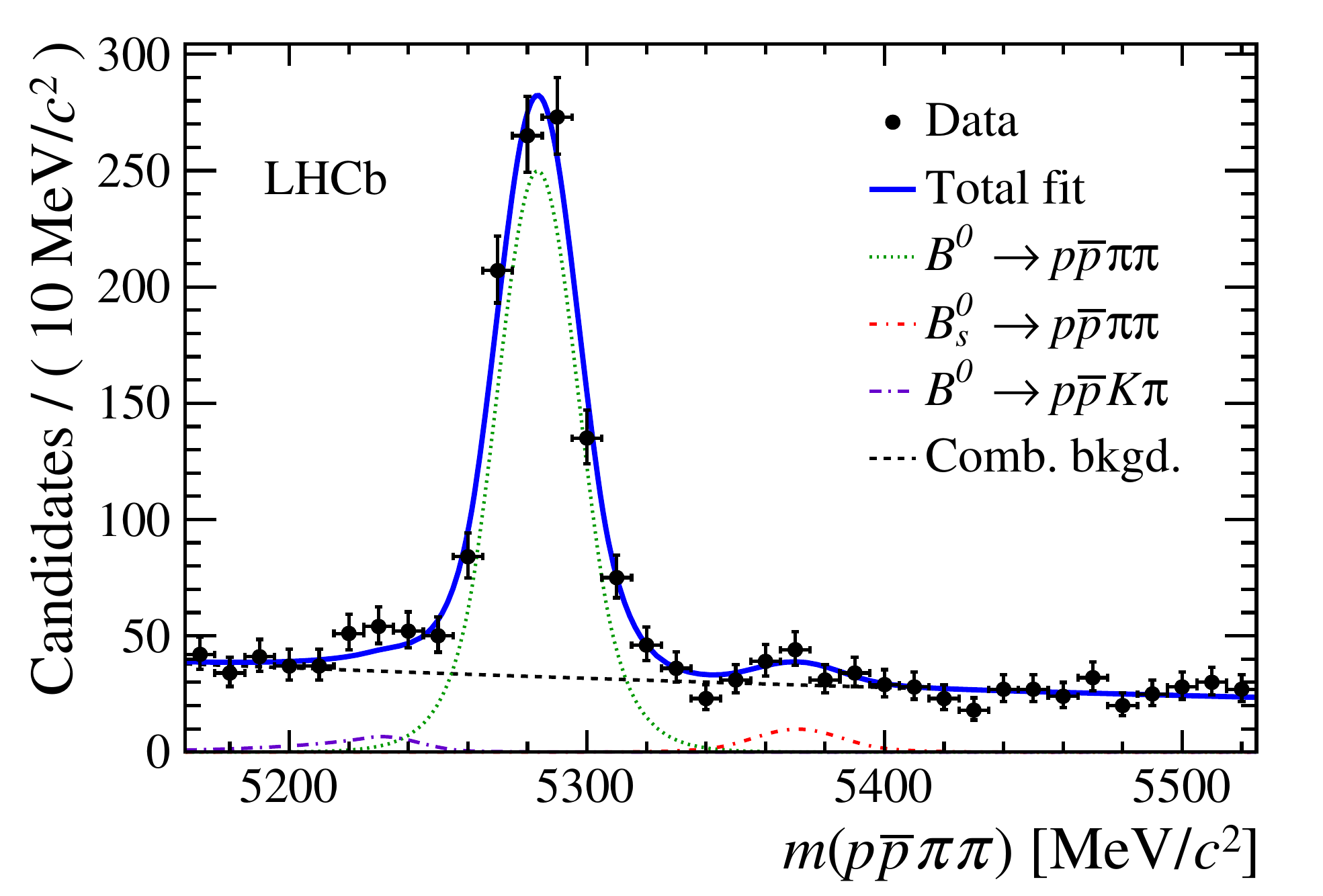}
  \includegraphics[width=0.49\textwidth]{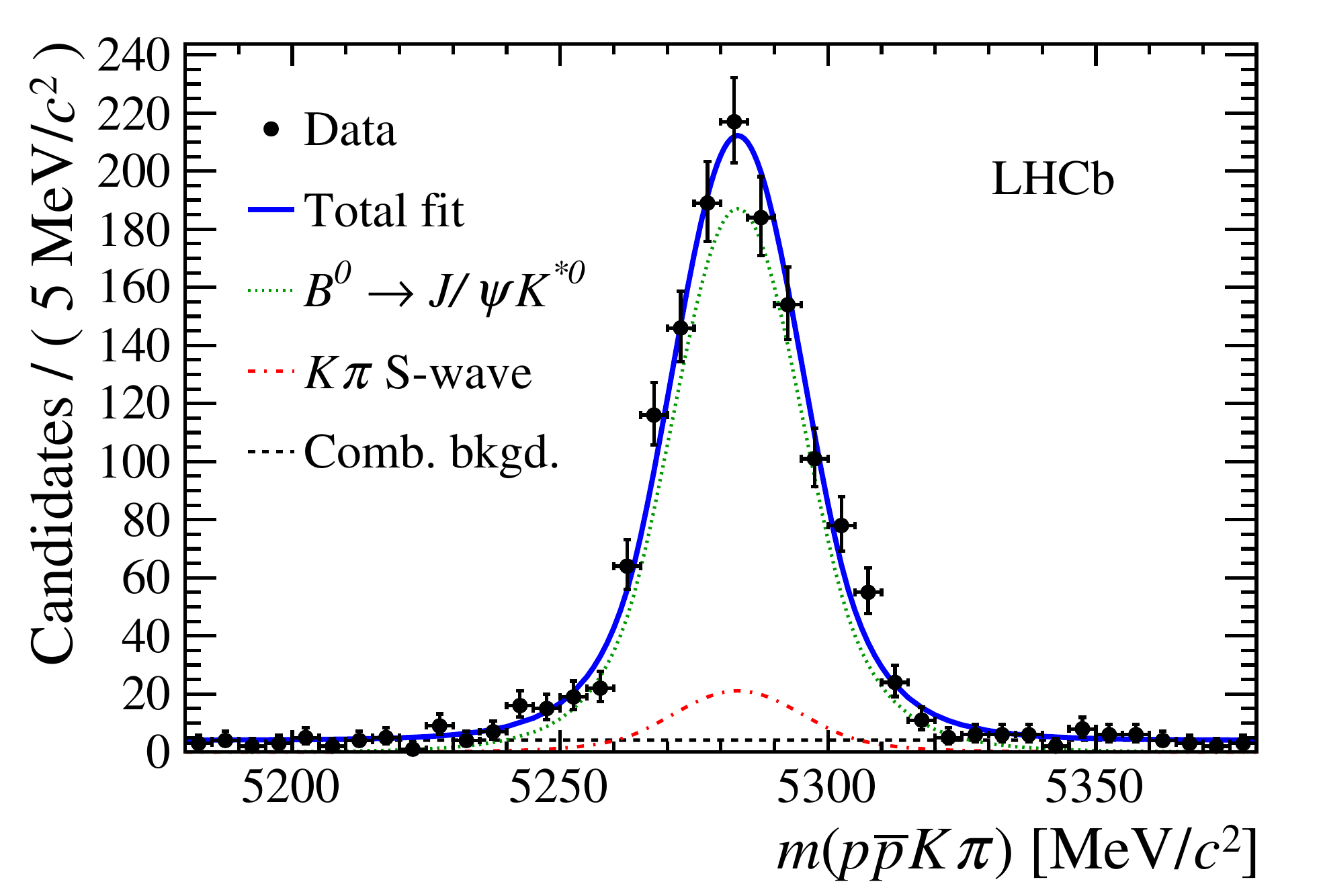}
  \caption{
    Invariant mass distributions for $\Bds$ candidates in the (top left)
    $\ppKK$, (top right) $\ppKPi$, (bottom left) $\ppPiPi$ final state
    and (bottom right) invariant mass distribution of \mbox{\sBToJPsiKst}
    in the $\ppKPi$ final state.
    The results of the fits are shown with blue solid lines.
    In the first three figures signals for $\Bd$ and $\Bs$ decays are
    shown, respectively, with green dotted and red dot-dashed lines, combinatorial
    backgrounds are shown with 
    black dashed lines and cross-feed backgrounds are shown with 
    violet dot-dashed lines. 
    In the bottom right figure the normalization signal is shown with a green dotted line, 
    the $\KPi$ S-wave component is displayed with a red dot-dashed line and the 
    combinatorial background with a black dashed line.
    } 
  \label{fig:fits}
\end{figure*}

The $\pphh$ invariant mass distributions with the results of the fit
overlaid are shown in Fig.~\ref{fig:fits} while the
signal
yields and the significances are collected in
Table~\ref{tab:results}. The
significance of each of the signal modes is determined from the change in
likelihood when the corresponding yield is fixed to zero, with
systematic uncertainties taken into account~\cite{Wilks:1938dza}. The
$\BsToppKPi$, $\BToppKK$ and $\BsToppPiPi$ modes are found to
have significances of $6.5$ standard deviations ($\sigma$), $4.1\,\sigma$
and $2.6\,\sigma$, respectively, while the
other signal modes have significances greater than $25\,\sigma$.

The branching fractions of the $\BTopphh$ decays are determined
relative to the visible branching fraction of the $\sBToJPsiKst$ decay using
\begin{equation}
\label{eq:BF}
\frac{\BF(\BTopphh)}{\BF_{\rm vis}(\sBToJPsiKst)} =
\frac{\mathcal{N^{\rm corr}}(\BTopphh)}{\mathcal{N^{\rm corr}}(\sBToJPsiKst)}\left(\times
  \frac{f_d}{f_s}\right)\,,
\end{equation}
where $f_s/f_d=0.259 \pm0.015$ (included only for the $\Bs$) is the ratio of $\bquark$ hadronization probabilities, $f_q$,
to the hadron $B_q$~\cite{fsfd},
and $\mathcal{N^{\rm corr}}$ denote efficiency-corrected fitted signal
yields.
The yields are obtained from the mass fits, while simulation is used to
evaluate the contribution to the efficiency from each stage
of the selection except for the effect of the PID criteria.  The latter is determined from 
calibration data samples of kinematically identified pions, kaons and protons
originating from the decays
$\decay{\Dstarp}{\Dz(\to\Km\pip)\pip}$, 
$\decay{\Lz}{\proton\pim}$ and
$\decay{\Lc}{\proton \Km \pip}$ 
and weighted according to the kinematics of the
signal particles~\cite{LHCb-DP-2012-003,Anderlini:2202412}.
For each final state the efficiencies are determined as a function of the position in 
phase space, and efficiency corrections for each
candidate are applied using the method of Ref.~\cite{LHCb-PAPER-2012-018}
to take the variation over the phase space into account. 
Explicitly, $\mathcal{N}^{\rm corr} = \sum_i \mathcal{W}_i/\epsilon_i$,
where the sum runs over the candidates in the fit, $\mathcal{W}_i$ is
the \sWeight for candidate $i$ determined with the \sPlot method~\cite{Pivk:2004ty} and
$\epsilon_i$ is the efficiency for the candidate $i$ which depends
only on its position in the five-dimensional phase space.
The visible branching fraction of the normalization mode,
defined as $\BF(\sBToJPsiKst) \times \BF(\mbox{\JPsiTopp}) \times\BF(\decay{K^*(892)^0}{\kaon^+\pion^-})$,
is $\BF_{\rm vis}(\sBToJPsiKst) = (1.68 \pm 0.12) \times 10^{-6}$,
where the $\sBToJPsiKst$ branching fraction is taken from
Ref.~\cite{Chilikin:2014bkk} and the others from Ref.~\cite{PDG2016}.

\begin{table*}[bt]
\caption{Fitted yields, signal yield significances and branching fractions computed using
  Eq.~(\ref{eq:BF}). The uncertainties on the yields are
  statistical only.
  The first uncertainty on each branching fraction is statistical, the second systematic, the third
  comes from the uncertainty on the branching fraction of the
  normalization mode and the fourth, where present, is due to the
  uncertainty on $f_d/f_s$.
\label{tab:results}}
\centering 
 \begin{tabular}{lr@{}lcl@{}l@{}l@{}l@{}l@{}l} 
  \toprule 
Decay  channel & \multicolumn{2}{c}{Yield $\mathcal{N}$} & Significance [$\sigma$]& \multicolumn{4}{c}{Branching fraction / $10^{-6}$} \\ 
 \midrule 
$B^0 \to p \overline{p} K K$ & $68$ & $\,\pm\, 17$  & 4.1 & $0.113$ & $\,\pm\,0.028$ & $\,\pm\,0.011$ & $\,\pm\,0.008$ \\ 
$B^0 \to p \overline{p} K \pi$ & $4155$ & $\,\pm\, 83$  & $> 25$ & $5.9$ & $\,\pm\,0.3$ & $\,\pm\,0.3$ & $\,\pm\,0.4$ \\ 
$B^0 \to p \overline{p} \pi \pi$ & $902$ & $\,\pm\, 35$  & $> 25$ & $2.7$ & $\,\pm\,0.1$ & $\,\pm\,0.1$ & $\,\pm\,0.2$ \\ 
$B^0_s \to p \overline{p} K K$ & $635$ & $\,\pm\, 32$  & $> 25$ & $4.2$ & $\,\pm\,0.3$ & $\,\pm\,0.2$ & $\,\pm\,0.3$ & $\,\pm\,0.2$ \\ 
$B^0_s \to p \overline{p} K \pi$ & $246$ & $\,\pm\, 39$  & 6.5 & $1.30$ & $\,\pm\,0.21$ & $\,\pm\,0.11$ & $\,\pm\,0.09$ & $\,\pm\,0.08$ \\ 
$B^0_s \to p \overline{p} \pi \pi$ & $39$ & $\,\pm\, 16$  & 2.6 & $0.41$ & $\,\pm\,0.17$ & $\,\pm\,0.04$ & $\,\pm\,0.03$ & $\,\pm\,0.02$ \\ 
$B^0 \to J/\psi K^{*}(892)^{0}$ & $1216$  & $\,\pm\, 45$ & -- & \multicolumn{4}{c}{--} \\ 
\bottomrule 
 \end{tabular}
\end{table*}

The branching fraction of each signal mode is reported in Table~\ref{tab:results}. The
significance for the $\BsToppPiPi$ mode is less than $3\,\sigma$; an
upper limit on its branching fraction is found~to~be
\begin{equation*}
\mathcal{B}(\BsToppPiPi) < 6.6\times 10^{-7}\, \mbox{at
$90\%$ confidence level},
\end{equation*}
by integrating the likelihood after multiplying by
a prior probability distribution that is uniform in the region of
positive branching fraction.
The values of the ratios of branching fractions between different
$\BTopphh$ decay modes are reported in Table~\ref{tab:ratioBRs}.

\begin{table}[bt]
\caption{Ratios of branching fractions among different $\BTopphh$ modes. The
  first uncertainty is statistical, the second systematic and the third, where present, comes from the
  uncertainty on $f_d/f_s$.
\label{tab:ratioBRs}}
\centering 
 \begin{tabular}{ll@{}l@{}l@{}l} 
  \toprule 
$\mathcal{B}(B^0 \to p \overline{p} K K)/\mathcal{B}(B^0 \to p \overline{p} K \pi)$ & $0.019$ & $\,\pm\,0.005$ & $\,\pm\,0.002$ \\ 
$\mathcal{B}(B^0 \to p \overline{p} \pi \pi)/\mathcal{B}(B^0 \to p \overline{p} K \pi)$ & $0.46$ & $\,\pm\,0.02$ & $\,\pm\,0.02$ \\ 
$\mathcal{B}(B^0_s \to p \overline{p} K \pi)/\mathcal{B}(B^0 \to p \overline{p} K \pi)$ & $0.22$ & $\,\pm\,0.04$ & $\,\pm\,0.02$ & $\,\pm\,0.01$ \\ 
$\mathcal{B}(B^0_s \to p \overline{p} K \pi)/\mathcal{B}(B^0_s \to p \overline{p} K K)$ & $0.31$ & $\,\pm\,0.05$ & $\,\pm\,0.02$ \\ 
\bottomrule 
 \end{tabular}
\end{table}

\begin{figure*}[tb]
  \centering
  \includegraphics[width=0.49\textwidth]{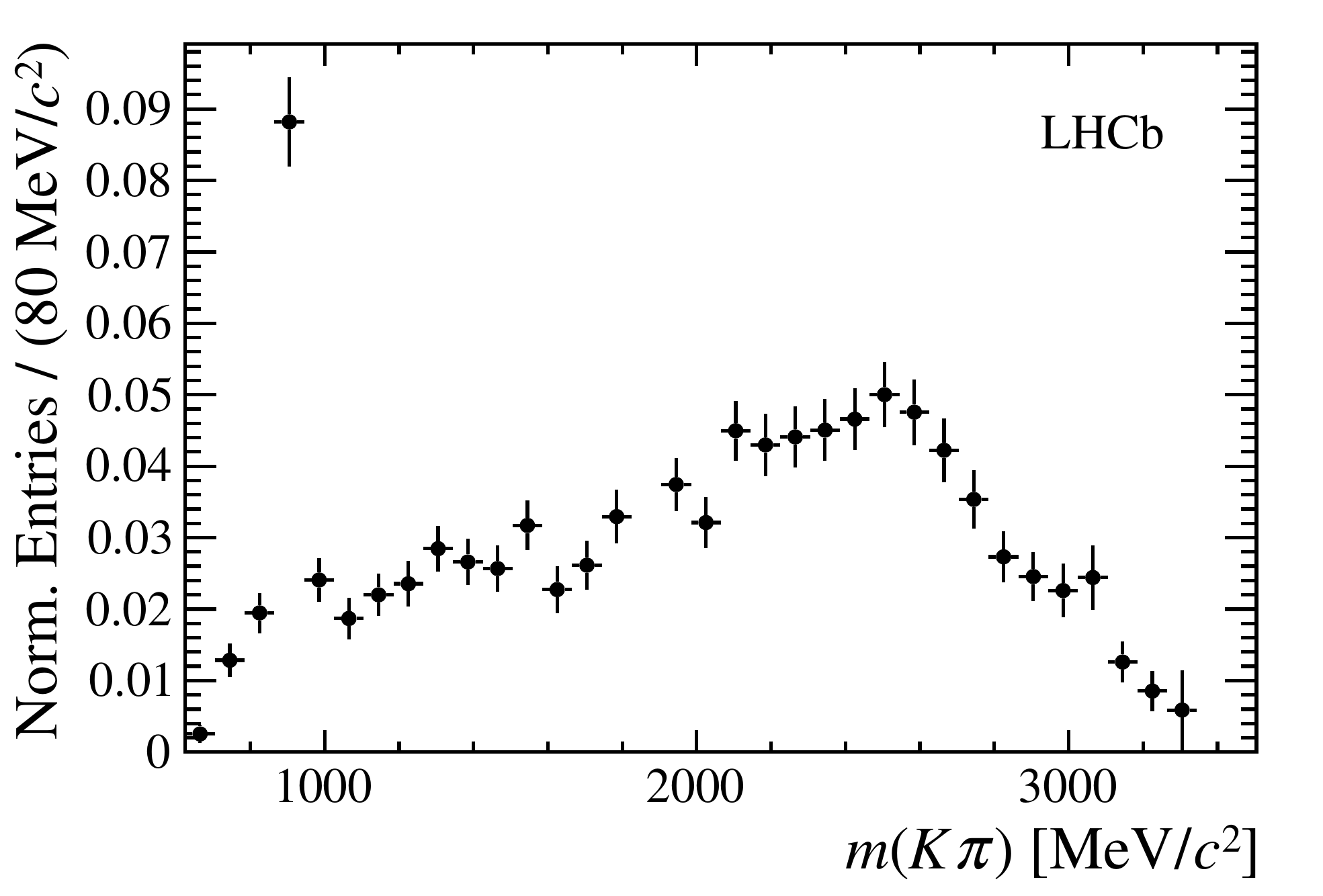}
  \includegraphics[width=0.49\textwidth]{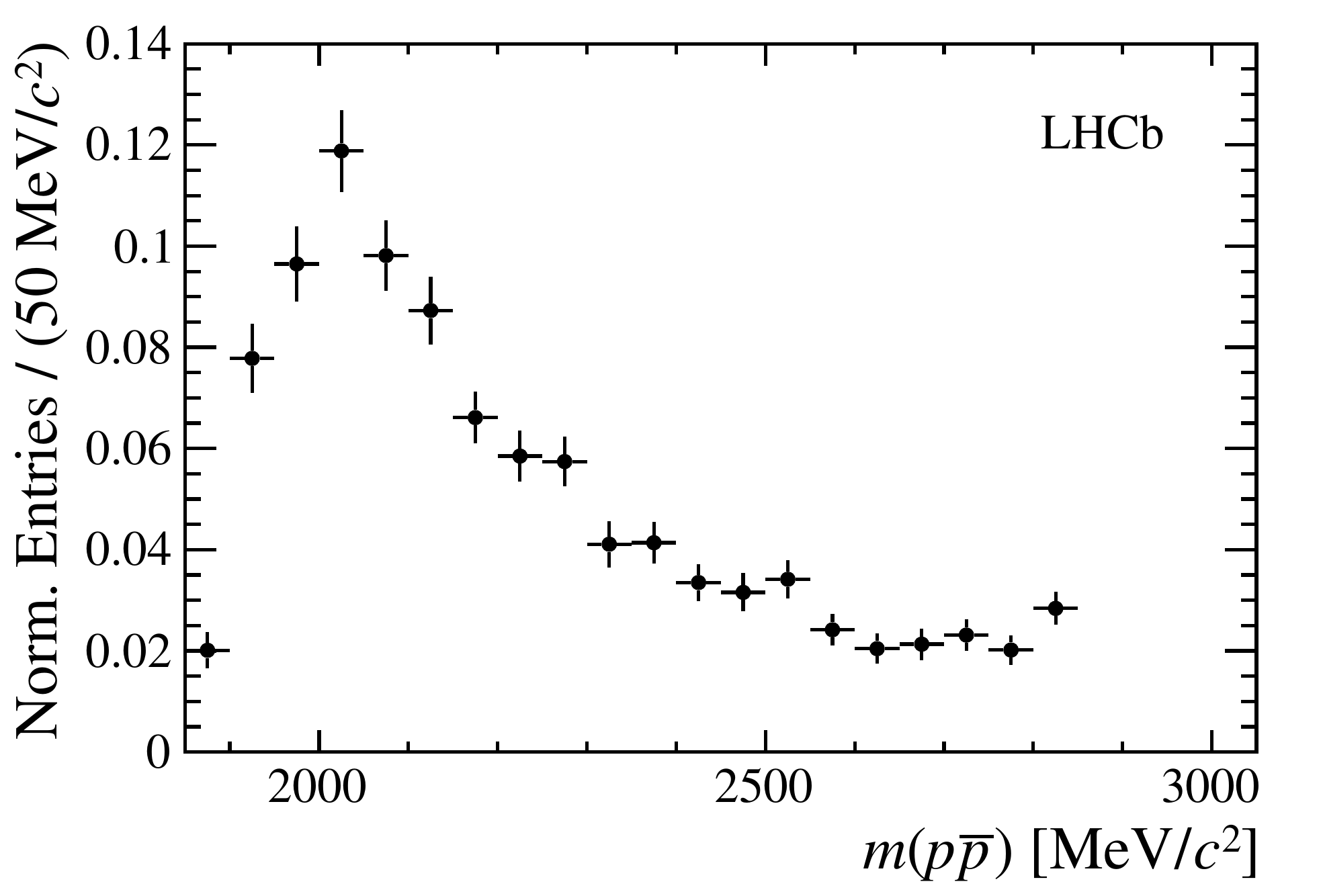}
  \includegraphics[width=0.49\textwidth]{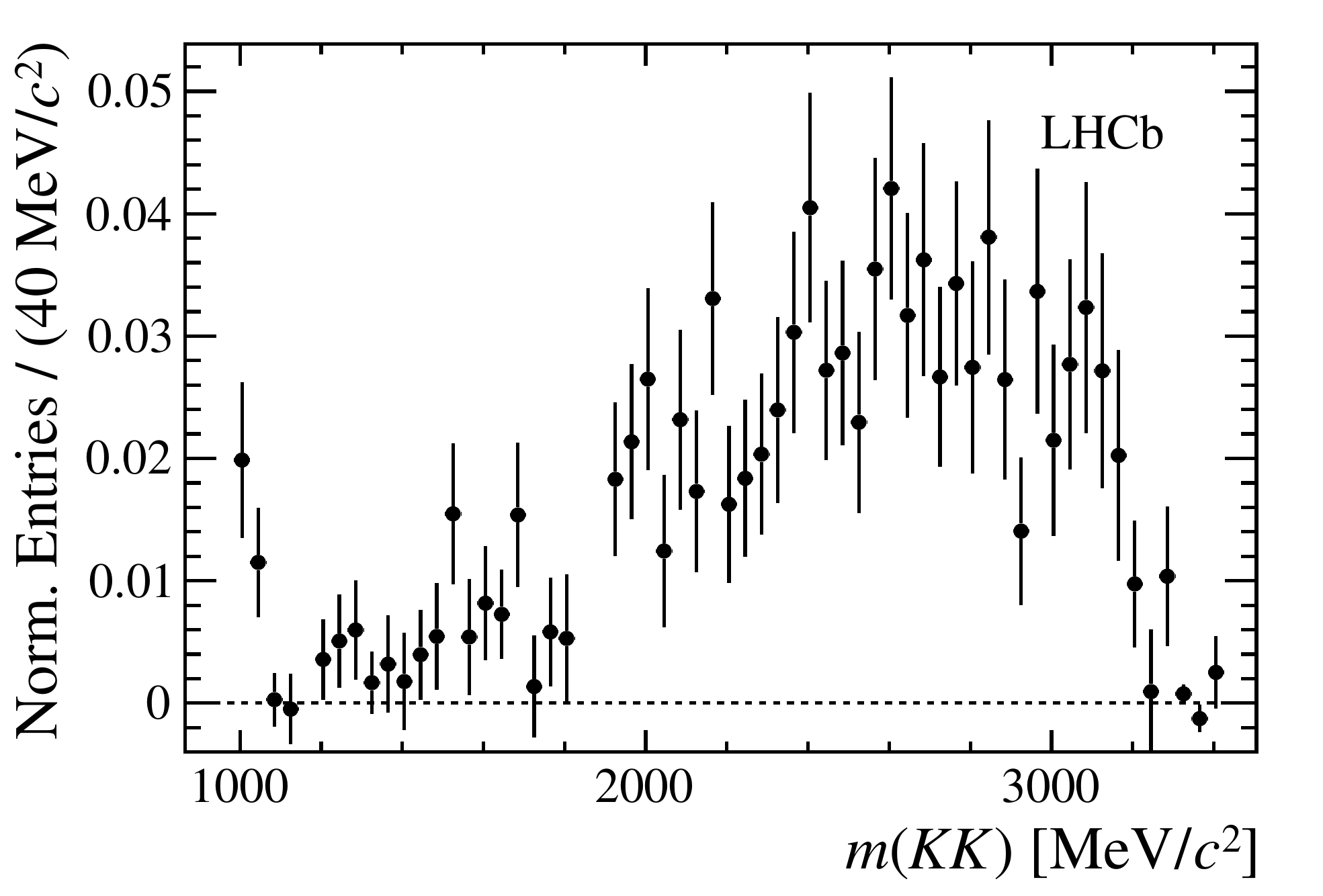}
  \includegraphics[width=0.49\textwidth]{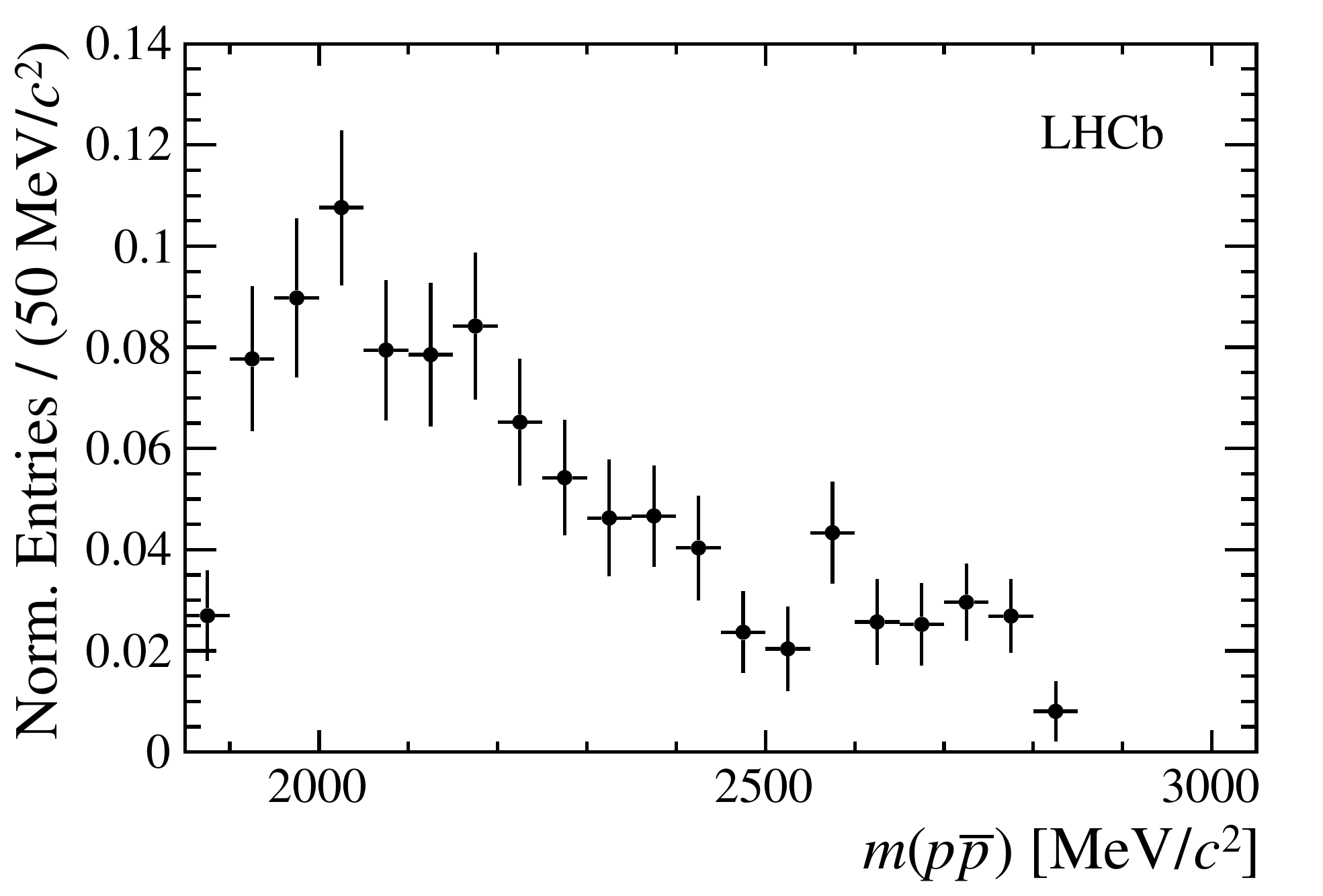}
  \includegraphics[width=0.49\textwidth]{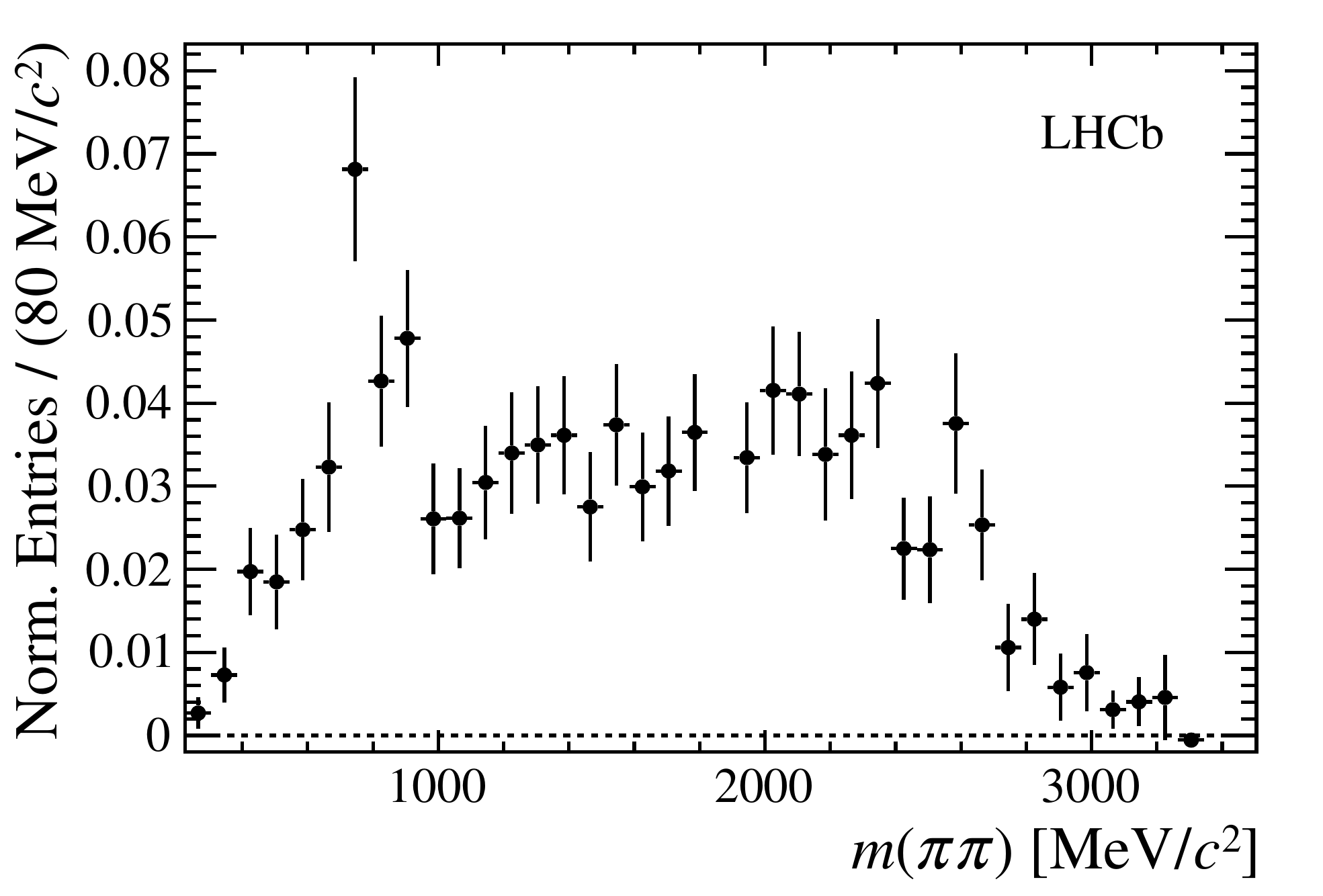}
  \includegraphics[width=0.49\textwidth]{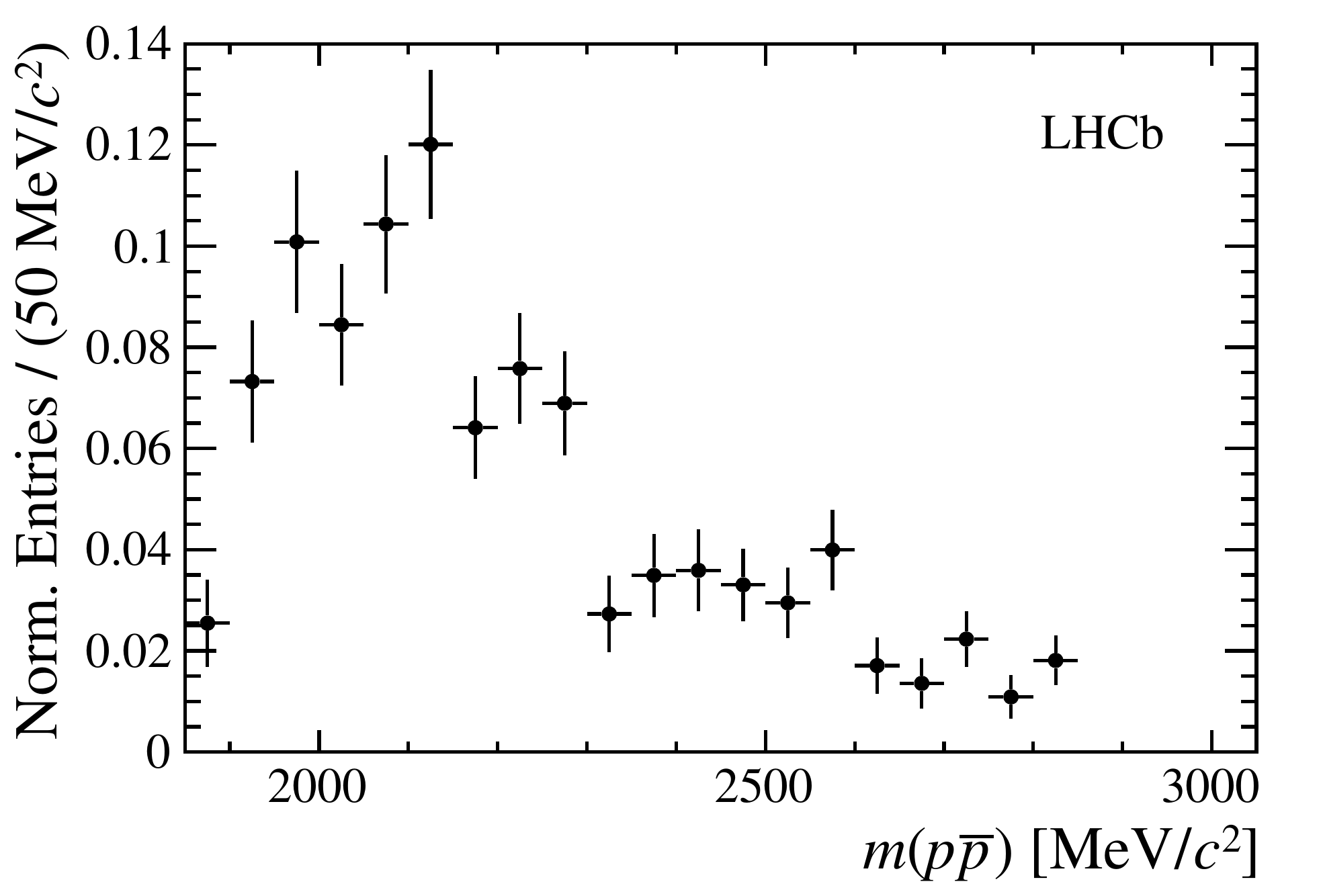}
  \caption{
    Efficiency-corrected and background-subtracted (left) $m(\hh)$ and (right) $m(\pp)$
    distributions from (top) $\BToppKPi$, (middle) $\BsToppKK$,
    and (bottom) $\BToppPiPi$ candidates.
    Events with entries in the charmonium or $D^0$ mass regions have been removed from the samples.
    All distributions are
    normalized to unity.
    } 
  \label{fig:substructures}
\end{figure*}

The signal distributions in $m(\hh)$ and $m(\pp)$ are obtained by subtracting the
background using the \sPlot technique~\cite{Pivk:2004ty}, with the $\Bds$ candidate invariant
mass as the discriminating variable. Per-candidate weights are applied to correct
for the variation of the selection efficiency over the phase space.
Figure~\ref{fig:substructures} shows the $\hh$ invariant mass distributions of
the $\BToppKPi$, $\BsToppKK$ and $\BToppPiPi$ decay modes. A peak from a vector meson
is identifiable in each mass spectrum, corresponding to a $K^*(892)^0$, a
$\Pphi(1020)$ and a $\rho(770)^0$  meson, respectively. The $\pp$ invariant mass distributions are also shown
for the same decay modes. An enhancement near threshold, typical in
baryonic \B decays~\cite{Hou:2000bz,Bevan:2014iga},
is clearly visible in each case. 
Detailed amplitude analyses of the $\BTopphh$ decays
will be of interest with larger samples.

The sources of systematic uncertainty on the absolute branching
fractions and on the ratios of branching fractions
arise from the fit model, the
knowledge of the efficiencies and, where appropriate, from the uncertainties on
the branching fraction of the normalization mode and on the
ratio of $b$-quark hadronization probabilities.
Pseudoexperiments are used to estimate the effect of using alternative shapes for
the fit components, or of including additional components in the fit.
In particular, the effect of adding other cross-feed
backgrounds, partially reconstructed backgrounds and components coming
from \Lb decays have been investigated. These are the dominant sources of systematic
uncertainty for the $\BToppKK$ and $\BsToppPiPi$ modes.
The effect of fixing the
yields of the cross-feed backgrounds based on the (mis)identification
probabilities is also assessed by varying these probabilities within
their uncertainties.
Intrinsic biases in the fitted yields are investigated with
pseudoexperiments and are found to be negligible. 
Uncertainties on the efficiencies arise due to the limited size of
the
simulation samples, the uncertainty on their evaluated distributions across
the phase space of the decays and 
from possible residual differences between data and simulation.
The unknown decay kinematics are the principal source
of systematic uncertainty for the $\BsToppKPi$ mode,
while for the $\BsToppKK$, $\BToppKPi$ and $\BToppPiPi$ modes the dominant source of systematic
uncertainty comes from the uncertainty on the efficiency of the
hardware stage of the trigger.
As the efficiencies depend on the signal decay-time distribution, 
the effect coming from the different lifetimes of the \Bs mass
eigenstates has been evaluated.
The systematic uncertainties due to the vetoes of charm hadrons are
also included.

In summary, a search for the four-body charmless baryonic decays $\BTopphh$
has been carried out by the \lhcb collaboration with a sample of proton-proton collision data
corresponding to an integrated luminosity of 3\invfb. 
First observations are obtained for the decays
$\BToppPiPi$, nonresonant $\BToppKPi$, $\BsToppKK$ and $\BsToppKPi$,
while first evidence is reported for the $\BToppKK$ mode and an upper
limit is set on the $\BsToppPiPi$ branching fraction.
In particular, four-body baryonic \Bs decays are observed for the
first time and a threshold enhancement in the baryon-antibaryon mass spectra
is confirmed for baryonic \Bs decays~\cite{LHCb-PAPER-2017-012}.

The \lhcb collaboration has recently published studies of \CP violation
with four-body $\LbTophhh$ decays studying triple-product correlations,
and presented first evidence for \CP violation in baryons~\cite{LHCb-PAPER-2016-030}.
The decays of \Bz and \Bs mesons to $\pphh$ final states reported in this paper
may be used in the future for similar studies of \CP violation in baryonic \B decays.

\section*{Acknowledgements}

\noindent We express our gratitude to our colleagues in the CERN
accelerator departments for the excellent performance of the LHC. We
thank the technical and administrative staff at the LHCb
institutes. We acknowledge support from CERN and from the national
agencies: CAPES, CNPq, FAPERJ and FINEP (Brazil); MOST and NSFC (China);
CNRS/IN2P3 (France); BMBF, DFG and MPG (Germany); INFN (Italy); 
NWO (The Netherlands); MNiSW and NCN (Poland); MEN/IFA (Romania); 
MinES and FASO (Russia); MinECo (Spain); SNSF and SER (Switzerland); 
NASU (Ukraine); STFC (United Kingdom); NSF (USA).
We acknowledge the computing resources that are provided by CERN, IN2P3 (France), KIT and DESY (Germany), INFN (Italy), SURF (The Netherlands), PIC (Spain), GridPP (United Kingdom), RRCKI and Yandex LLC (Russia), CSCS (Switzerland), IFIN-HH (Romania), CBPF (Brazil), PL-GRID (Poland) and OSC (USA). We are indebted to the communities behind the multiple open 
source software packages on which we depend.
Individual groups or members have received support from AvH Foundation (Germany),
EPLANET, Marie Sk\l{}odowska-Curie Actions and ERC (European Union), 
Conseil G\'{e}n\'{e}ral de Haute-Savoie, Labex ENIGMASS and OCEVU, 
R\'{e}gion Auvergne (France), RFBR and Yandex LLC (Russia), GVA, XuntaGal and GENCAT (Spain), Herchel Smith Fund, The Royal Society, Royal Commission for the Exhibition of 1851 and the Leverhulme Trust (United Kingdom).

\clearpage
\FloatBarrier

\clearpage
\FloatBarrier
\addcontentsline{toc}{section}{References}
\setboolean{inbibliography}{true}
\ifx\mcitethebibliography\mciteundefinedmacro
\PackageError{LHCb.bst}{mciteplus.sty has not been loaded}
{This bibstyle requires the use of the mciteplus package.}\fi
\providecommand{\href}[2]{#2}

\newpage
\centerline{\large\bf LHCb collaboration}
\begin{flushleft}
\small
R.~Aaij$^{40}$,
B.~Adeva$^{39}$,
M.~Adinolfi$^{48}$,
Z.~Ajaltouni$^{5}$,
S.~Akar$^{59}$,
J.~Albrecht$^{10}$,
F.~Alessio$^{40}$,
M.~Alexander$^{53}$,
S.~Ali$^{43}$,
G.~Alkhazov$^{31}$
% ...

{\footnotesize \it
% ...
$ ^{5}$Clermont Universit{\'e}, Universit{\'e} Blaise Pascal, CNRS/IN2P3, LPC, Clermont-Ferrand, France\\
% ...
$ ^{10}$Fakult{\"a}t Physik, Technische Universit{\"a}t Dortmund, Dortmund, Germany\\
% ...
$ ^{31}$Petersburg Nuclear Physics Institute (PNPI), Gatchina, Russia\\
% ...
$ ^{39}$Universidad de Santiago de Compostela, Santiago de Compostela, Spain\\
$ ^{40}$European Organization for Nuclear Research (CERN), Geneva, Switzerland\\
% ...
$ ^{43}$Nikhef National Institute for Subatomic Physics, Amsterdam, The Netherlands\\
% ...
$ ^{48}$H.H. Wills Physics Laboratory, University of Bristol, Bristol, United Kingdom\\
% ...
$ ^{53}$School of Physics and Astronomy, University of Glasgow, Glasgow, United Kingdom\\
% ...
$ ^{59}$University of Cincinnati, Cincinnati, OH, United States\\
% ...
}
\end{flushleft}

\end{document}